# An Autonomous GIS Agent Framework for Geospatial Data Retrieval

Huan Ning, Zhenlong Li[*], Temitope Akinboyewa, M. Naser Lessani

Geoinformation and Big Data Research Laboratory, Department of Geography
The Pennsylvania State University, University Park, USA
[*]zhenlong@psu.edu

**Abstract**: Powered by the emerging large language models (LLMs), autonomous geographic information systems (GIS) agents have the potential to accomplish spatial analyses and cartographic tasks. However, a research gap exists to support fully autonomous GIS agents: how to enable agents to discover and download the necessary data for geospatial analyses. This study proposes an autonomous GIS agent framework capable of retrieving required geospatial data by generating, executing, and debugging programs. The framework utilizes the LLM as the decision-maker, selects the appropriate data source (s) from a pre-defined source list, and fetches the data from the chosen source. Each data source has a handbook that records the metadata and technical details for data retrieval. The proposed framework is designed in a plug-and-play style to ensure flexibility and extensibility. Human users or autonomous data scrawlers can add new data sources by adding new handbooks. We developed a prototype agent based on the framework, released as a QGIS plugin (*GeoData Retrieve Agent*) and a Python program. Experiment results demonstrate its capability of retrieving data from various sources including OpenStreetMap, administrative boundaries and demographic data from the US Census Bureau, satellite basemaps from ESRI World Imagery, global digital elevation model (DEM) from OpenTopography.org, weather data from a commercial provider, the COVID-19 cases from the NYTimes GitHub. Our study is among the first attempts to develop an autonomous geospatial data retrieval agent.

## 1   Introduction

In recent years, large language models (LLMs) have drawn tremendous attention from researchers. These models are trained on massive text corpora and are able to reason and respond to questions (prompts) in natural language. Serving as reasoning cores, LLMs are able to make decisions and assist analysis for various applications, such as robotics (Sun et al., 2024) and clinical diagnosis (Hong et al., 2024). LLM can also power geographic information systems (GIS), enabling GIS tools to perform spatial analysis tasks autonomously. Researchers in the GIScience community have attempted to implement such autonomous GIS agents, aiming to automate spatial analysis (Li & Ning, 2023; Zhang et al., 2023), cartography (Zhang et al., 2024), and disaster management (Akinboyewa et al., 2024; Hao et al., 2024). The main purpose of those autonomous agents is to promote productivity and democratize GIS technologies by reducing human intervention.

Before conducting any spatial analyses, GIS agents need to access essential geospatial datasets, such as study area boundaries, points of interest (POIs), imagery, and attributes of geographic entities. Li & Ning (2023) highlighted that online geospatial data discovery and filtering are necessary capabilities of autonomous GIS. Their experiments showed that LLM is able to obtain data from REST (Representational State Transfer) APIs with brief documentation about endpoints, parameters, and format of the returned data. In their study, however, most experimental datasets were provided by local file paths or URLs. Their work on autonomous GIS did not propose solutions or mechanisms for geospatial data retrieval.

Geospatial data can be sourced from a variety of online and local repositories. For example, in various research contexts, OpenStreetMap, governmental agencies, and satellite imagery repositories are widely used as open-access data sources. Automatically retrieving data from these extensive data sources can enhance GIS agents' autonomy, potentially enabling fully autonomous research. Zhang et al. (2023) proposed an autonomous GIS framework, GeoGPT, to collect, process, and analyze geospatial data. GeoGPT is able to download POIs and road networks from OpenStreetMap and remote sensing images from Geospatial Data Cloud. However, the authors did not provide details on the implementation of the data fetching module and its expanding mechanism. The work by Feng et al. (2023) on GeoQAMap exhibited the idea of querying data from an online Wikidata and Query Service data source. GeoQAMap first adopted an LLM (GPT-3.5, OpenAI, 2024) to convert the spatial questions on geo-entities to SPARQL queries, and then retrieved related geo-entities from Wikidata. The authors suggested that GeoQAMap could potentially be applied to other data sources, such as OpenStreetMap and CityGML, but did not provide further investigation or a general framework for data retrieval. Similar to GeoQAMap, Staniek et al. (2023) also investigated the feasibility of data fetching from OpenStreetMap by applying fine-tuned T5 models (Raffel et al., 2020) and GPT-4 (OpenAI, 2024) to convert questions in natural language to queries in OverpassQL, a query language for OpenStreetMap data. GeoForge (Ageospatial, 2024) is a web application powered by GPT that can request, process, analyze, and display satellite imagery, building footprints, and administrative boundaries. Its goal is to democratize geospatial information and analysis. However, this closed-sourced application is still in development, with limited data sources and no support for customized and local data sources.

While researchers have demonstrated the feasibility of LLM-powered agents to handle geospatial data queries using natural language, universal mechanisms and implementations are rarely detailed in the literature. Autonomous geospatial data retrieval faces several critical challenges, including 1) discovering relevant data from various online sources. While search engines like Google can retrieve data based on keywords, selecting and assessing appropriate geospatial data requires comprehensive strategies. This process is both time-consuming and intelligence-intensive; 2) selecting the appropriate data sources by interpreting data requests made in natural language or specific requirements, which may not always clearly designate the data source. The autonomous agent needs to understand such requests and determine which data source should be used; and 3) fetching data from specific sources involves varying technical methods. Some data sources provide API (Application Programming Interface) access, downloading URLs, and sometimes require API keys or passwords. Autonomous agents need to be flexible, knowledgeable, and "smart" to handle the parameters in APIs or URLs correctly.

To address this gap, we designed an autonomous GIS agent framework for geospatial data retrieval, aiming to address the second and third challenges. The challenge of online data source discovery will be explored in future research as it requires intensive research into web search and analysis. The framework retrieves requested data by generating, debugging, and executing programs based on natural language queries. It is expandable and flexible, utilizing handbooks for each data source that provide necessary metadata and technical information for data fetching. To demonstrate the concept and feasibility of the framework, we developed a prototype agent including six data sources: OpenStreetMap, US Census data, weather data, ESRI (Environmental Systems Research Institute, Inc.) satellite images, COVID-19 data, and worldwide digital elevation models (DEMs). New data sources can be added to the agent by adding new handbooks in a plug-and-play manner. For over 70 data request tests, the agent obtained the correct data with a high chance of 80% - 90%, demonstrating its potential for automating geospatial data retrieval for autonomous GIS.

We implemented the prototype agent in two formats: a QGIS plugin named *Autonomous GIS - GeoData Retrieve Agent* and an interactive Python program in Jupyter Notebook. The QGIS plugin allows GIS users to download geospatial data without coding and use the downloaded data directly in a GIS environment. Meanwhile, developers can use the Python program to download data with greater flexibility and customization, such as automating geoprocessing tasks or making the agent collaborate with other programs or agents.

## 2  Methodology

Typically, GIS analysts find and download geospatial data in three steps: 1) collect potential data sources and their metadata, such as extent, resolution, date, and attributes. This procedure may involve using search engines or consulting colleagues; 2) determine which data source is most suitable for the analysis; 3) fetch the data based on the technical requirements set by the data providers, such as accessing URLs, using registered accounts, understanding data structures, and handling data chunking. In other words, human analysts are required to follow technical handbooks to retrieve the data.

Correspondingly, a geospatial data retrieval agent is expected to conduct these three steps autonomously. Since a GIS analyst is often specified in one or several domains, in this study we began with a manually created GIS data source list and handbooks and focus on the automation of the latter two steps: determining the applicable data source from an existing list and fetching data from the selected source. This section introduces the proposed framework, LLM-Find, and explains its implementation details through a proof-of-concept prototype.

### 2.1  The framework

The framework comprises two major components: a data source index and a handbook inventory (Figure 1). The data source index provides brief introductions to various data sources, enabling LLMs to select the most appropriate source based on the data request. The handbook inventory contains technical handbooks detailing how to fetch data from each source, such as accessing entries and data formats. The aim is to provide sufficient information to reduce uncertainty for the agents. Additionally, two supplementary components, *authentication* and *codebase*, are used to manage source authorization and provide example code. The framework is designed to be straightforward and extendable to accommodate various data sources across different application scenarios. This design allows both humans and other autonomous agents to update or add new data sources in a plug-and-play manner.

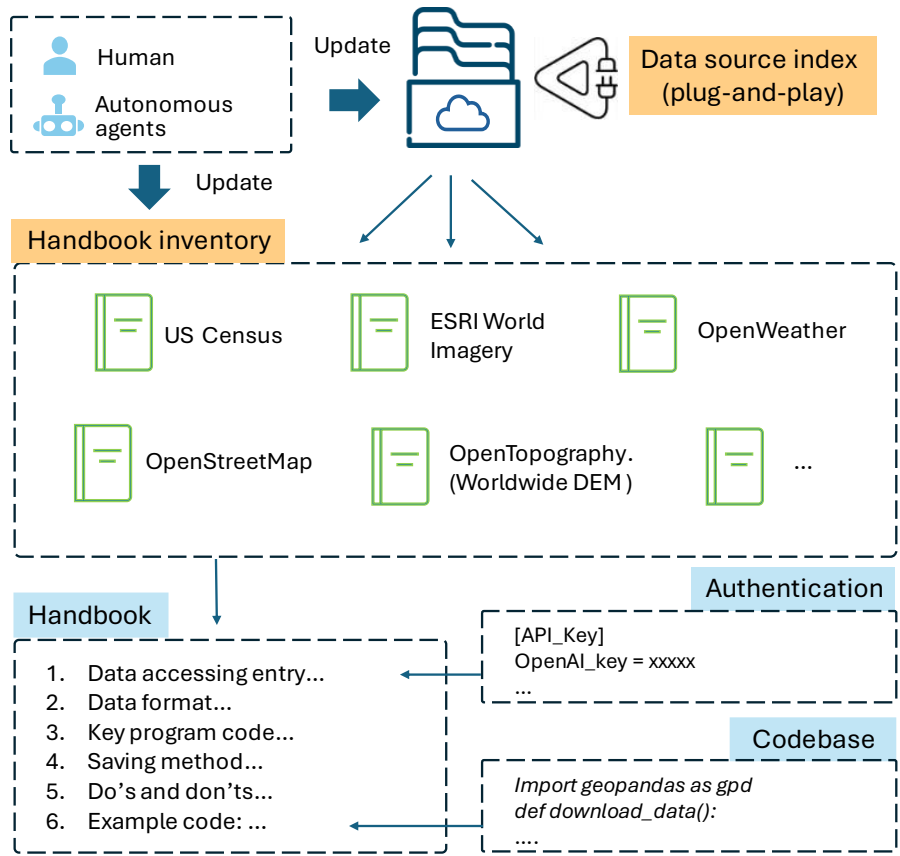

Figure 1. The architecture of the autonomous geospatial data retrieval framework

### 2.1.1 Data source index

The data source index includes the source names and concise descriptions, informing the agent about the data sources it can use. A sample of the data source index is shown in Box 1, including six data sources. Note that the data source index should be brief and informative since it needs to be embedded in the prompt. Because the descriptive names may be lengthy and unsuitable for use as data source identifiers (IDs), we defined short aliases for them to serve as IDs.

Box 1. An example of the data source index.

> 1. OpenStreetMap. You can download the administrative boundaries, street networks, points of interest (POIs) from OpenStreetMap.
>
> 2. US Census Bureau boundary. It provides the US administrative boundaries (nation, state, county, tract, and block group level, as well as metropolitan statistical areas.
>
> 3. US Census Bureau demography. It provides the demographic and socio-economic data, such as population, gender, income, education and race.
>
> 4. OpenWeather data. It provides historical, current, and forecast weather data. The historical data can be back to 2023-08. API limits: [Hourly forecast: 4 days, Daily forecast: 16 days, 3 hour forecast: 5 days]
>
> 5. ESRI World Imagery (for Export). It is a web map service, providing satellite image tiles. You can download tiles and mosaic them into a large image.
>
> 6. US COVID-19 data by New York Times. Cumulative counts of COVID-19 cases and deaths in the United States, at the state and county level, over time from 2020-01-21 to 2023-03-23.
>
> 7. OpenTopography. You can download global digital elevation model (DEM) data using API; the resolution ranges from 15m to 1000m, such as SRTM GL3 (global 90m), and GL1 (global 30m). The DEM source list from this API contains: SRTMGL3, SRTMGL1, SRTMGL1_E, AW3D30, AW3D30, SRTM15Plus, NASADEM, COP30, COP30, EU_DTM, GEDI_L3, GEBCOIceTopo, GEBCOSubIceTopo.

### 2.1.2 Handbook inventory

Each handbook in the inventory provides the necessary technical information for LLMs to retrieve data from its associated source. Although LLMs have been trained on massive texts, they still lack the detailed knowledge required to retrieve GIS data from various sources. Handbooks essentially fill the knowledge gaps in LLMs for particular geospatial data sources. They might cover data location, entry endpoints, format, and attributes. For sources requiring authentication (e.g., API keys) or specialized code to download and process data, the handbook provides this information. Since the framework generates programs to retrieve data, we found that providing it with a verified program template can significantly increase its code generation accuracy and reliability. These template programs are integrated into prompts and managed within a supplementary component, *codebase*, to ensure better maintenance and organization of these templates. In addition, the framework includes a dedicated component to manage authentication for security concerns. Each handbook serves a single data source and is maintained individually. The framework can add, update, and remove data sources by modifying the data source index and handbook inventory, making it plug-and-play. While humans can manually create and maintain these data sources, autonomous agents can also perform this task by crawling and assessing data sources and automatically generating handbooks with minimal human intervention.

Handbooks vary among data sources. For instance, OpenStreetMap provides the Overpass Query Language for API data retrieval, while the US Census Bureau offers download links for administrative boundaries. There may also be multiple methods or sources to retrieve the data. For example, GIS analysts can obtain OpenStreetMap data using the official website's export function, the Overpass API, or Python packages like OSMnx, each with its advantages and disadvantages. Moreover, both OpenStreetMap and the US Census Bureau provide US administrative boundaries, so determining the better fit for a specific spatial analysis is crucial. Therefore, the agent needs to be provided with necessary information via handbooks to ensure reliable data fetching. In this study, we manually created handbooks based on intensive experiments to ensure agents can correctly understand the data request. Box 2 lists a few guidelines in the handbook for fetching OpenStreetMap data, including API endpoints and conditions for using specific Python packages.

Box 2. Parts of the handbook for retrieving OpenStreetMap data.

> 1. In the requested area is given in an English name, you need to use `['name:en'='XX']` to filter the place in Overpass queries, otherwise you will get empty results. The `name` tag in OpenStreetMap usually is in the location language.
> 2. If you need to download the administrative boundary of a place from OpenStreetMap, please use a Python package named 'OSMnx' by this code line: `ox.geocode_to_gdf (query, which_result=None, by_osmid=False, buffer_dist=None)`. This method is fast.
> 3. If you need to download POIs, you may use the Overpass API, which is faster than OSMnx library. Code example is: `area['SO3166-2'='US-PA']->.searchArea;(nwr[amenity='hospital'](area.searchArea););out center;`
> 4. If you need to download polylines, you may use the Overpass API, which is faster than OSMnx library.
> 5. If you need to use a boundary to filter feature in GeoPandas, this is the code: `gpd.sjoin(gdf, boundary, how='inner', op='within')`.
> …
> This is a program for your reference, note that you can improve it: ….
> import geopandas as gpd
> def download_data():
>     overpass_url = "https://overpass-api.de/api/interpreter"
>     overpass_query = """
> …

## 2.2 The proof-of-concept prototype agent

Based on the framework, we developed a proof-of-concept prototype agent that utilizes GPT-4o as the decision core to select the data source and generate Python programs to retrieve geospatial data. Figure 2 demonstrates the workflow of the agent. First, the agent receives the data request described in natural language and sends it along with the data source index to LLM to select the most appropriate source. The data request can be straightforward, such as "Download the 2021 population by race for each county in the US and save the downloaded data to D:\data.csv". Upon receiving the LLM's reply, the agent extracts the selected data source and retrieves the handbook associated with the selected source, which includes necessary authentications and template programs. Appendix 1 shows an example prompt for selecting the data source.

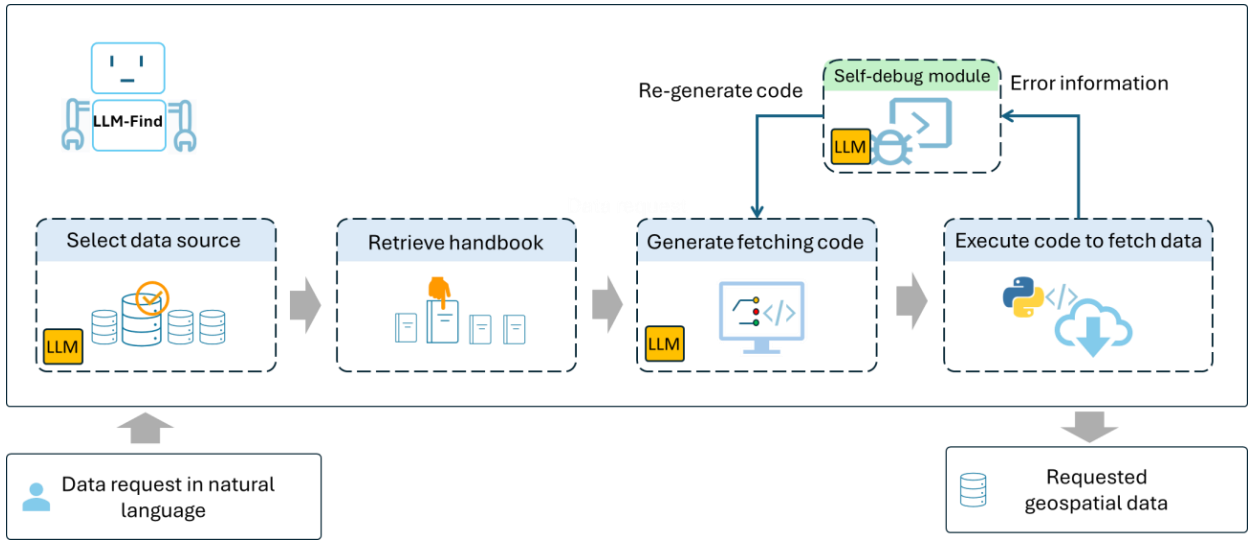

Figure. 2 Workflow of the protopye agent based on the proposed framework

With the data source identified and the handbook retrieved, the agent will then ask the LLM to generate Python code to fetch data. The handbook that contains the technical requirements will be sent with the code generation request. The agent employs a restricted program structure to ensure the code runs in a Python environment. For example, the LLM is instructed to generate the program using a pre-defined structure, such as placing all code within a "download_data()" function rather than arbitrary names or the "__main__" function. The agent incorporated such information and requirements into the prompt to get the data downloading code (a prompt example can be found in Appendix 2). Finally, the agent executes the generated code to fetch the data.

Recognizing that the LLM's spatial programming ability is limited due to insufficient training data and the complexity of the data-fetching programs, it is rare for LLMs (and humans) to write a bug-free program on the first attempt. Code debugging is a common and necessary step in real-world programming. Therefore, we implemented a self-debug module for the agent to correct the buggy code according to the error information. The self-debug module sends the necessary information to LLM and asks it to correct the code. The sent information includes the data request, handbook, generated program, occurred error, and debugging requirements. The re-generated code will be executed again to fetch the data. This execution-debugging process often runs multiple times until the generated program is error-free or the iteration times exceed the pre-defined limitation (which is set to 10 times in the prototype agent).

We developed the prototype agent in two forms: a QGIS plugin (*Autonomous GIS - GeoData Retrieve Agent*) and an interactive Python program in Jupyter Notebook. The QGIS plugin can help users download geospatial data without coding and use the data directly in the QGIS environment. QGIS is a free, open-source, and cross-platform GIS software, supporting geospatial data visualization and analysis. Around 2 million users install QGIS worldwide (Gispo, 2021). The plugin can help these users to retrieve and load data using one simple sentence in natural language, aiming to improve their work efficiency. The interactive Python program allows GIS programmers to download data with greater flexibility and customization, such as automating their geoprocessing tasks or making the agent collaborate with other programs or agents. Figure 3 shows the QGIS plugin with downloaded data. The plugin is docked on the right side of QGIS map canvas, showing the generated code and agent running information. Users can type their data requests, set the output directory, and press the "Send Request" button to retrieve the data. The downloaded data will be automatically loaded into QGIS for further analysis.

## 3 Case Studies

The prototype agent currently supports six data sources, covering data types of vector, raster, and attribute. This section presents the results obtained by the agent when requesting data from each of these sources. We extensively tested the agent across various scenarios, resulting in a code repository with over 70 test cases. We encourage readers to explore these cases or to test their own data requests using either the QGIS plugin or the Python program.

### 3.1 *OpenStreetMap*

OpenStreetMap is a collaborative project that creates a free, editable map of the world, built by volunteers who collect and contribute geographic data (Haklay & Weber, 2008). It offers a vast repository of spatial data, including roads, trails, points of interest, and administrative boundaries, making it a valuable resource for various GIS applications. The data is structured using nodes, ways, and relations, which can be queried through the Overpass API. Our experiments showed that downloading data from OpenStreetMap was challenging, including generating correct Overpass queries and parsing results. The complexity of the OpenStreetMap data structure and tags increases the difficulties of fetching and saving the data for autonomous agents. For example, GPT-4o struggled to convert OpenStreetMap's *'relation'* data structure into multi-polygons with multi-holes. However, we significantly improved the success rate by providing a template program in the handbook. Figure 3 shows the results of downloading city points, river polylines, and state administrative polygons of Nigeria using the QGIS plugin. Attributes were also downloaded as '*tags'* in OpenStreetMap.

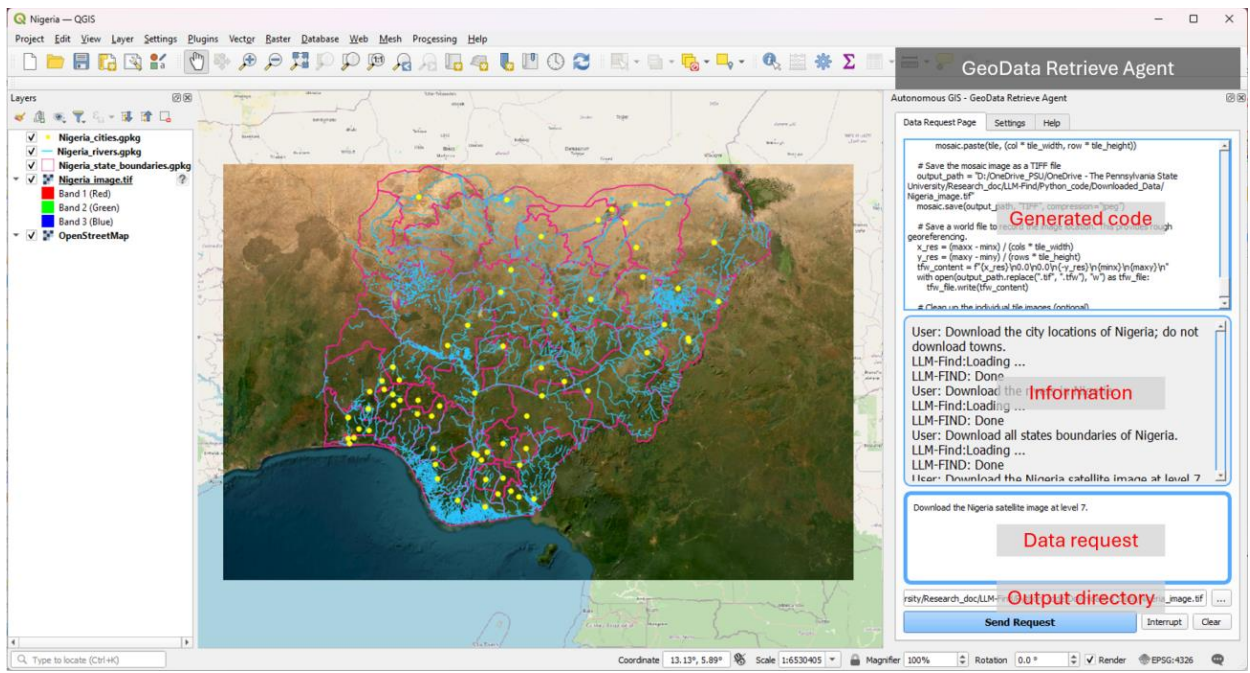

Figure 3. The QGIS plugin of the GeoData Retrieve Agent and the downloaded geospatial data of Nigeria, including cities (point), rivers (polyline), and state boundaries (polygon) from OpenStreetMap. The image basemap was downloaded from ESRI World Imagery using the plugin. Note that there are four individual data requests to retrieve the shown data (e.g., "Download the rivers in Nigeria.").

## 3.2 US Census data

US Census data includes comprehensive demographic information about the population of the US, including variables such as population size, age distribution, income levels, education, and housing characteristics at varying geographic levels (e.g., blockgroup, census tract, county, and state). The agent handbook for the Census data includes the data access API information along with a template program to support the agent in selecting the needed variables from more than 27,000 available variables. Additionally, the US Census Bureau's TIGER data (Topologically Integrated Geographic Encoding and Referencing) provides detailed geographic information about various administrative boundaries in the US, such as states, counties, and census tracts. These datasets are essential for geographic analysis and mapping. The handbook includes the boundary file URL structure for the agent to select the region and level, such as tract, county, or state based on the request.

Figure 4 presents the county boundaries of Pennsylvania downloaded with the QGIS plugin and the population for each county in Pennsylvania in 2022. Figure 5 shows the results of the another two requests: "Download the population over 25 years old and the population with a college degree or higher at the state level of USA for 2012 and 2022" and "Download the state boundary of the contigous United States". Multiple variables across multiple years can be retrieved in one request as shown in Figure 5. However, we observed that the agent may query the wrong variable combinations at times when the data involves multiple variables. For instance, when asking for the total senior population, it may only query the male or female population.

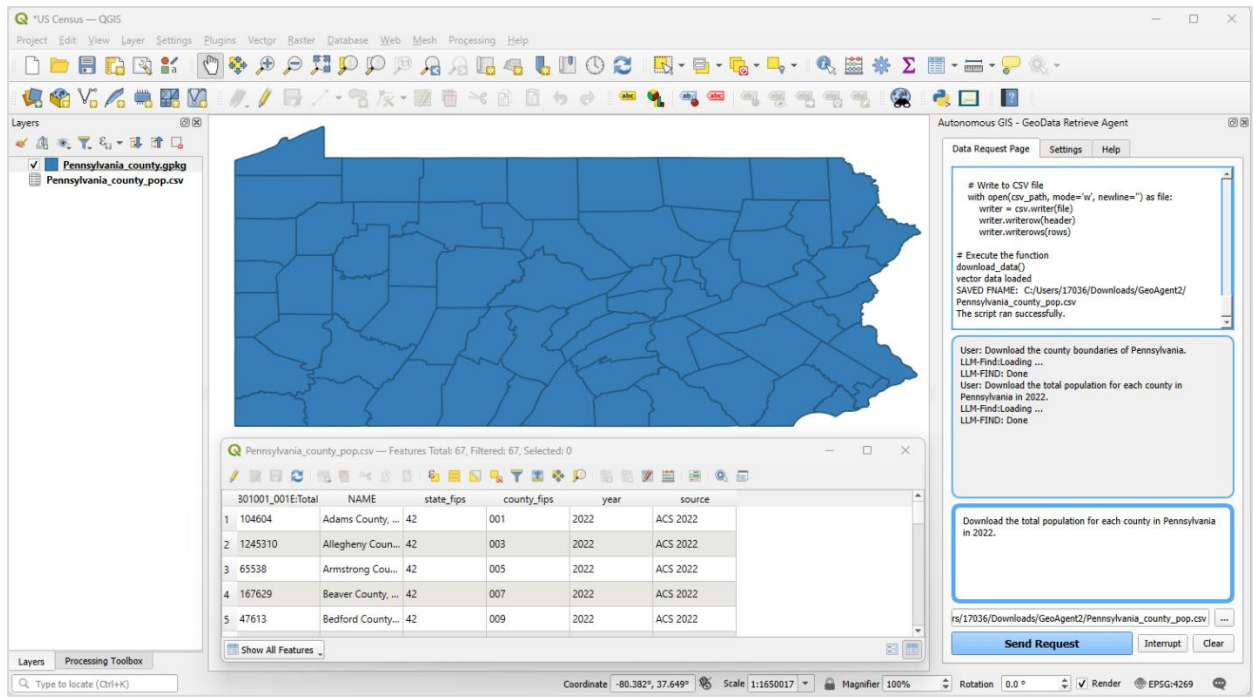

**Figure 4.** County boundaries of Pennsylvania downloaded with the QGIS plugin (request: *Download the county boundaries of Pennsylvania.*) and the population for each county in Pennsylvania in 2022. (request: *Download the total population for each county in Pennsylvania in 2022*).

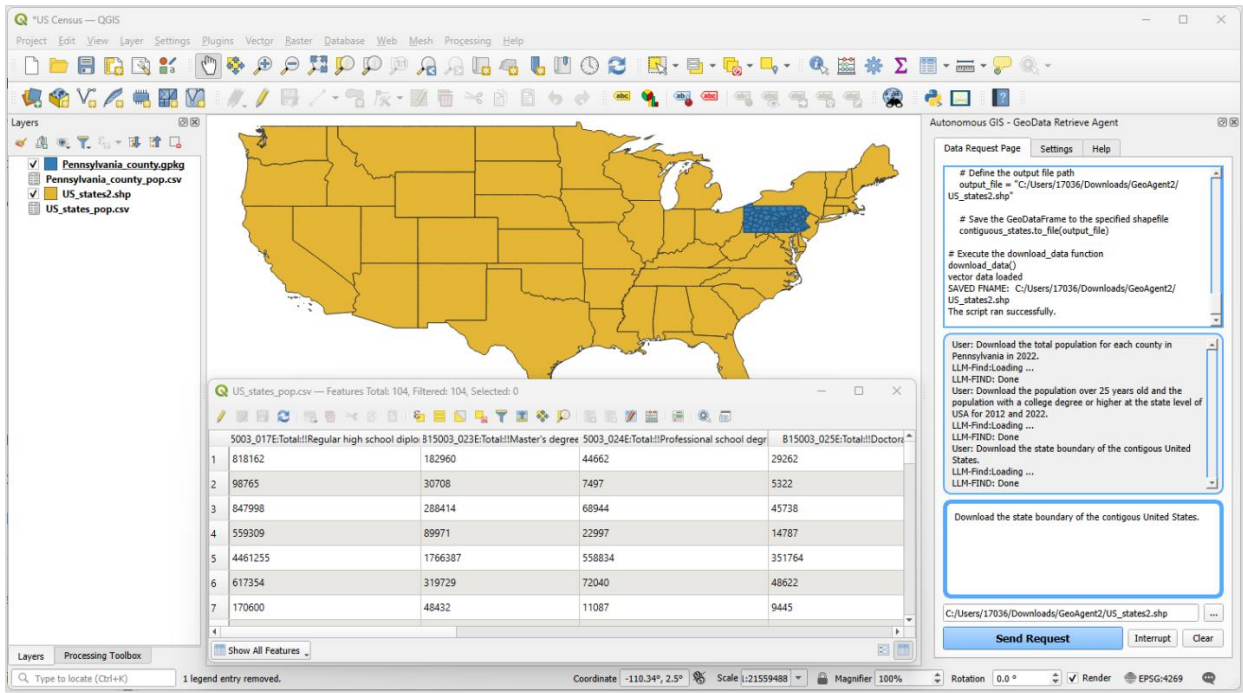

**Figure 5.** The downloaded *state boundary of the contigous United States* and *the population over 25 years old and the population with a college degree or higher at the state level of USA for 2012 and 2022*.

## 3.3 Satellite imagery

Satellite imagery is an essential type of Earth observation data used in both natural and social science research. However, retrieving satellite images is not trivial as researchers typically need a background in satellite sensors, accounts with data providers, and expertise in processing large volumes of image data. We equipped the protopye agent with access to an image tile web service from ESRI World Imagery, enabling it to download images based on place names and bounding boxes. The agent adoptes the geocoding service provided by nominatim.org to retrieve the boundary from the given place name to identify the image extent.

Figures 6 shows the downloaded satellite images of the FAST Telescope in China (the top map), Yellowstone National Park (Map 1), a region defined by a bounding box (Map 2), Qingdao City in China (Map 3). Another example is shown in Figure 3, featuring the downloaded image of Nigeria. These examples, covering various geographic extents and resolution levels, demonstrate the feasibility of downloading satellite images with autonomous agents. New data sources along with corresponding handbooks can be added to the agent to enhance its capability in fetching other remote sensing images.

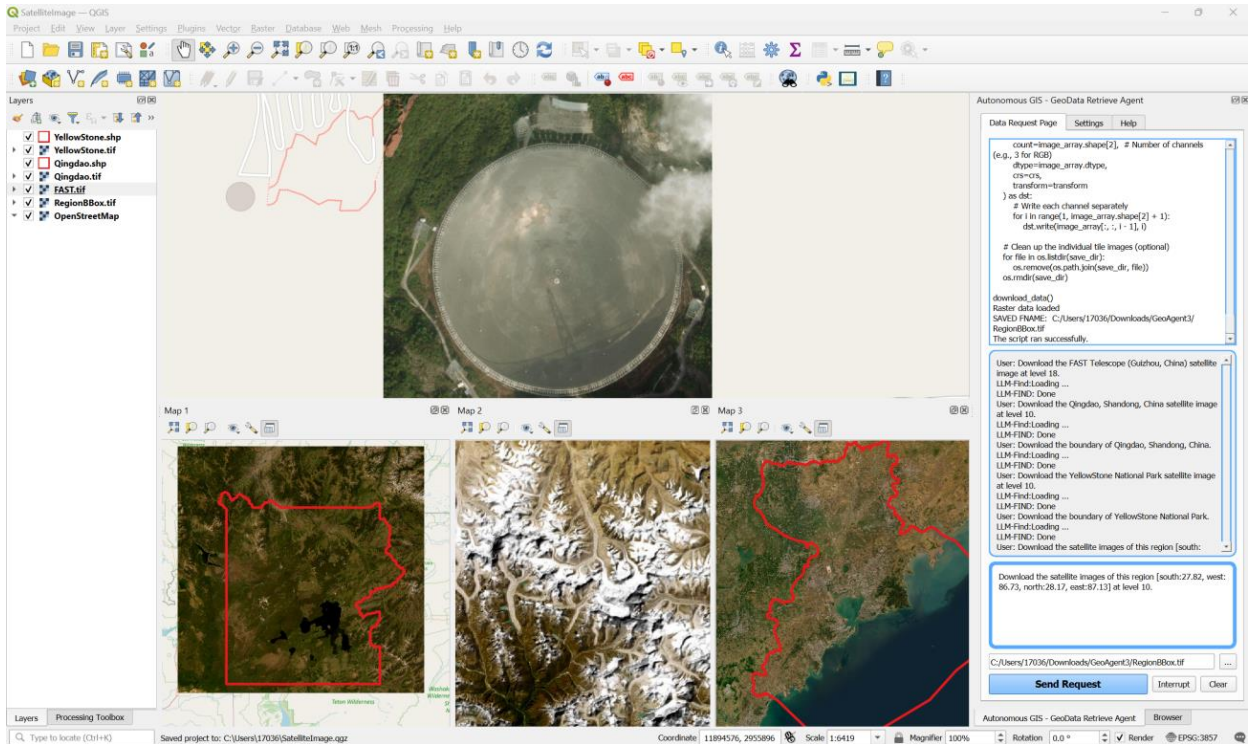

**Figure 6**. The downloaded satellite images. **Top map**: the FAST Telescope in China (request: *Download the FAST Telescope (Guizhou, China) satellite image at level 18.*). **Map 1**: Yellowstone National Park (request: *Download the Yellowstone National Park satellite image at level 10.*); **Map 2**: a region defined by a bounding box (request: *Download the satellite images of this region [south:27.82, west:86.73, north:28.17, east:87.13] at level 10.*); **Map 3**: Qingdao City in China (request: *Download the Qingdao, Shandong, China satellite image at level 10*). Note that the red boundaries in Map 1 and Map 3 were downloaded using the plugin by two separate requests based on the place names*: Download the boundary of Yellowstone National Park* and *Download the boundary of Qingdao, Shandong, China.*

## 3.4   *Digital elevation model (DEM)*

DEM is a 3D representation of a terrain surface in raster format, and it is an important environmental factor in various applications, such as ecology research and infrastructure planning. OpenTopography.org provides massive DEM products worldwide. The agent incorporates OpenTopography's global DEM products of 30m and 90m resolution and is able to retrieve DEM data based on a place name or bounding box. The top map of Figure 7 shows the downloaded DEM of Puerto Rico overlapped with three other data layers retrieved from OpenStreetMap including county boundaries, major rivers, and hospitals. In addition, the county level population and median household income of Puerto Rico in 2020 were also downloaded from the US Census using the plugin. The bottom map of Figure 7 shows the 30-meter resolution DEM of Chongqing, China downloaded with another request.

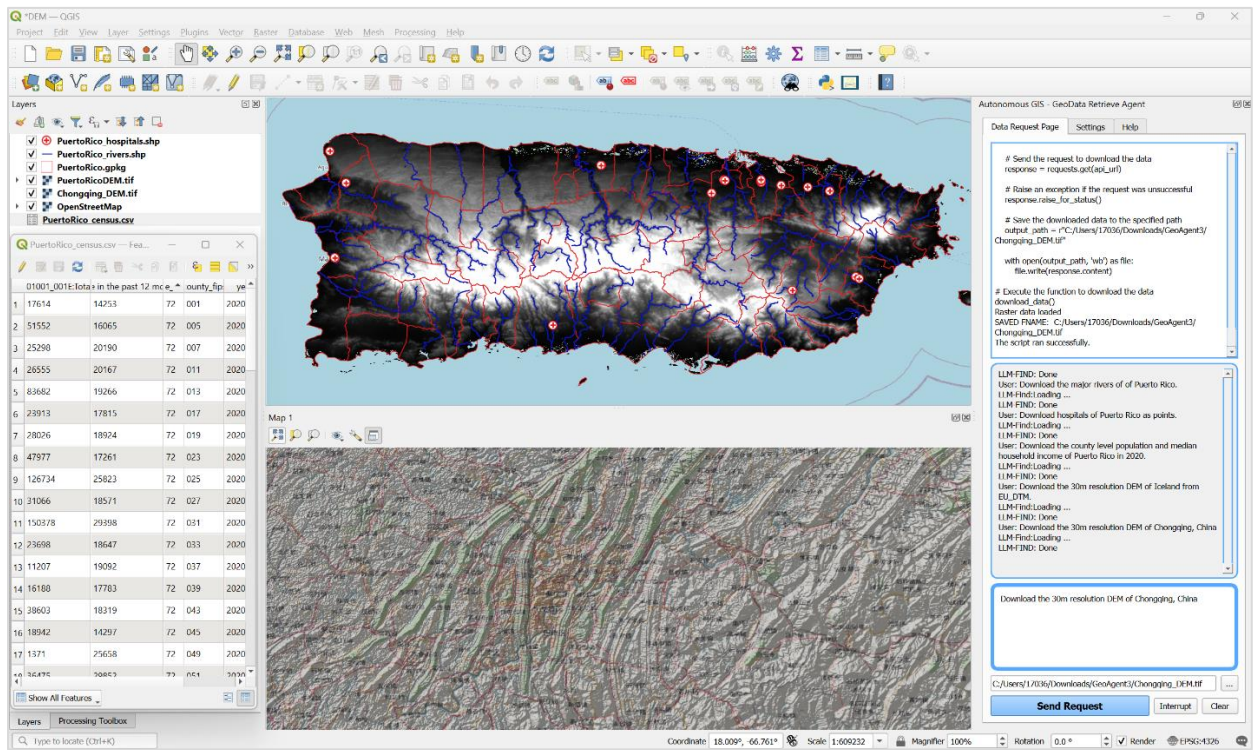

**Figure 7**. **Top map**: DEM of Puerto Rico overlapped with three other data layers retrieved from OpenStreetMap including county boundaries, major rivers, and hospitals. (DEM request: *Download the 90-meter resolution DEM data of Puerto Rico*.) **Bottom map**: DEM of Chongqing, China (Request: *Download the 30m resolution DEM of Chongqing, China)*. The DEM is displayed as Hillshade on top of the OpenStreetMap basemap with a transparency of 50%. **Left table**: the county level population and median household income of Puerto Rico in 2020 downloaded from the US Census using the plugin.

## 3.5   Weather data

We integrated the weather data API from OpenWeather (openweathermap.org) into the protopye agent. This API accepts location and period parameters to return historical, current, and forecast weather data. Note that the GPT-4o cannot get the current time on its own, so we added explicit prompts in the handbook to specify the period (historical, current, or forecast) or the current date and time to select appropriate API endpoints and parameters. Figure 8 shows an example of downloading historical weather data for Yulin City in Guangxi, China during May 2024 using the Jupyter Notebook-based agent. Examples of fetching current weather data can be found in Figure 9. Since weather data is inherently spatiotemporal, this case study demonstrates that the agent can effectively manage both spatial and temporal aspects simultaneously, which is crucial in spatial analysis.

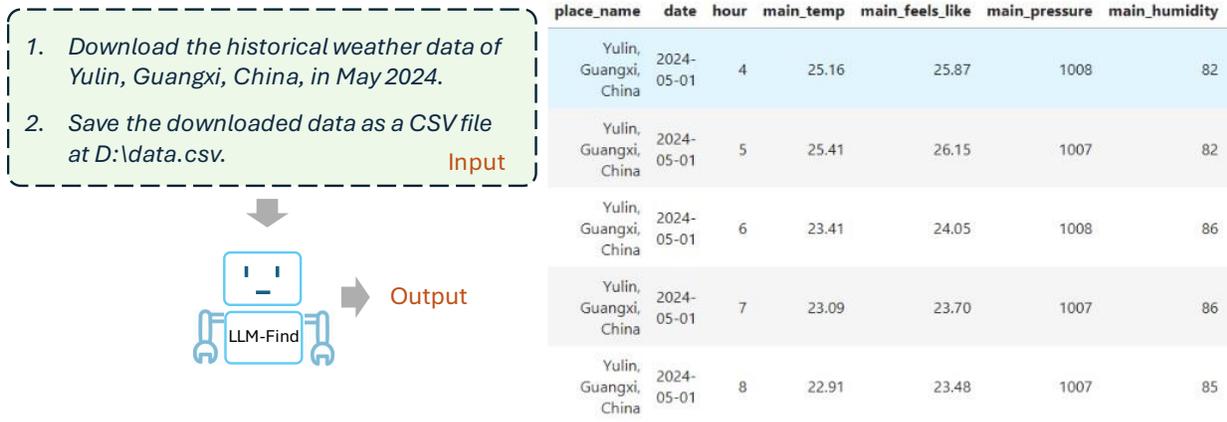

**Figure 8.** Download the historical weather data for Yilin, Guangxi, China.

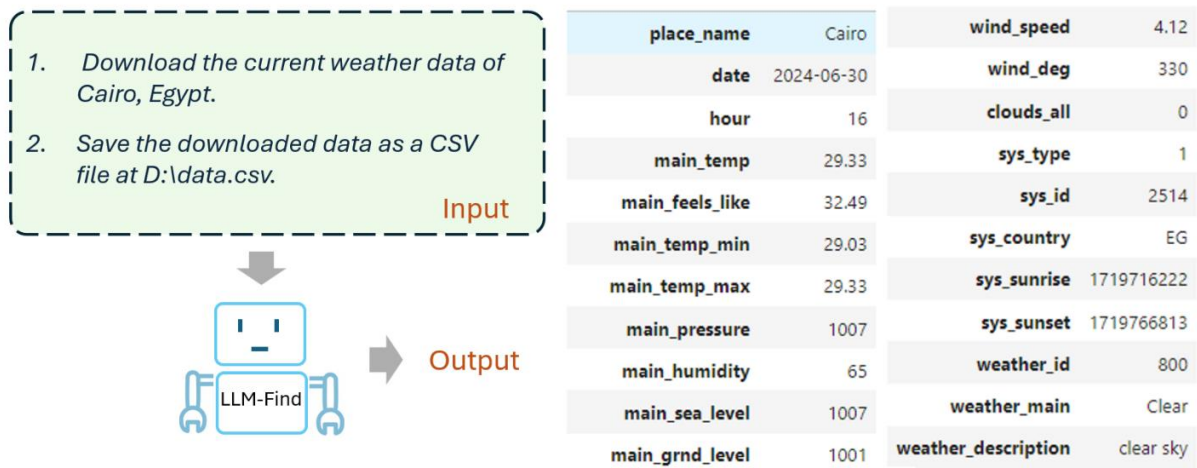

**Figure 9.** Download current weather data for Cairo, Egypt.

### 3.6 COVID-19 data

The New York Times provides a comprehensive dataset on COVID-19 accumulative cases, tracking the spread and impact of the virus across various regions (New York Times, 2023). This dataset includes daily updates on case numbers, deaths, and other relevant metrics during the pandamic, making it a valuable resource for public health analysis and research. Similar to the US Census Bureau administrative boundaries, this dataset can be accessed via URLs. We provided the agent with a handbook containing data URLs, data structures, and date ranges. As illustrated in Figure 10, the agent can download COVID-19 accumulative case data based on the requested geographic level and time period. This study case demonstrates the agent's ability to handle data sources without formal APIs.

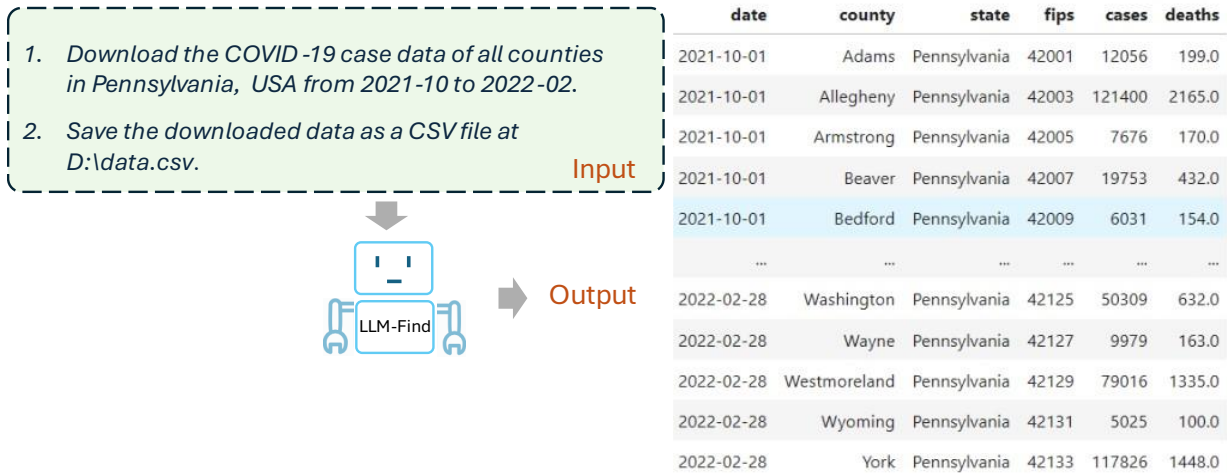

**Figure 10.** Download accumulative COVID-19 cases for counties in the US from October 2021 to February 2022.

## 4 Discussion

The proposed autonomous agent framework, LLM-Find, demonstrated the feasibility of retrieving geospatial data by simple natural language command input without human intervention, indicating that such data retrieval agents can provide necessary data to other agents for further analysis. This research reflects the advocation of autonomous GIS (Li & Ning, 2023), which envisions that the next generation of GIS could conduct spatial analysis with less or without human intervention.

The framework supports a plug-and-play system and handbook hubs as demonstrated by the prototype agent and case studies. First, this design is highly extendable: adding new data sources only requires updating the index and creating the corresponding handbook, with no impact on existing sources, which simplifies maintenance. The geospatial community can establish handbook hubs to share well-written and validated handbooks. Data providers should consider releasing data not only for human users but also for autonomous agents. Thus, the traditional geospatial metadata format or standard may need to be updated to meet the technical requirements of such agents. Second, the proposed framework provides the flexibility to allow another agent to crawl online geospatial data sources, update the index and create handbooks autonomously. For example, LLM-Find can be connected to an *autonomous data discovery agent* powered by LLMs for Internet search, webpage analysis, geospatial data assessment, and handbook generation.

Despite the promising results, the current implementation of the framework has several limitations. One major limitation is that it only supports text handbooks, which are unsuitable for some applications. For example, many geospatial data sources, particularly remote sensing images, have particular boundaries represented by vector files or raster maps. The framework needs further development to support multi-modal handbooks. Another limitation is the inability to support long handbooks due to the prompt length restrictions of LLMs. Some data sources require extensive materials to fetch data correctly. For example, the US Census data provides more than 27,000 fine-grained variables for the American Community Survey 5-year survey (US Census Bureau, 2023). Retrieving the correct variable combinations from this large list is challenging. A possible solution is to feed all variables and their descriptions into the handbook, leveraging LLMs that support large input tokens, such as Gemini 1.5 (Google, 2024). However, computational time and cost may pose obstacles. Therefore, approaches such as Retrieval Augmented Generation (RAG) (Lewis et al., 2020) should be explored to support long and comprehensive handbooks.

Additionally, a data understanding and assessment module is necessary for the geospatial data retrieval agent. Human GIS analysts often choose between similar datasets or make compromises when encountering unsuitable datasets. For instance, they need to decide to use 1:5,000,000 or 1:50,000 scale boundaries for country-level analysis or select low-resolution terrain data when high-resolution data is unavailable. Such decisions require experience, detailed knowledge of existing data, a profound understanding of application contexts, and creative problem-solving skills. Future research could evaluate whether autonomous agents can perform these tasks as well as or better than humans.

## 5 Conclusion

The proposed autonomous GIS agent framework has demonstrated the feasibility of retrieving geospatial data through simple natural language commands without human intervention. This capability indicates that such data retrieval agents can effectively provide necessary data to other agents for further analysis, supporting the vision of autonomous GIS. The framework's plug-and-play design and handbook hubs make it extendable and easy to maintain. Adding new data sources requires only updating the index and creating corresponding handbooks without affecting existing sources. This feature allows the geospatial community to share well-written and validated handbooks seamlessly. Moreover, the framework allows another agent to autonomously crawl online geospatial data sources, update the index, and create handbooks. This integration can enhance the autonomous data discovery process.

Based on the framework, we developed a proof-of-concept data retrieval agent with two implementation: a QGIS plugin named *GeoData Retrieve Agent* and a Python program in the Jupyter Notebook environment. The case studies demonstrated that the prototype agent could select the appropriate data source, load the associated handbook, generate and execute a Python program to retrieve the requested geospatial data. Our study is among the first to develop an autonomous geospatial data retrieval agent. Future research should focus on enhancing autonomous online data discovery, assessment, and handbook generation to further advance autonomous GIScience research.

**Source code availability statement:** The source code of the LLM-Find agent (Python program with Jupyter Notebook) along with over 70 data request examples can be found at https://github.com/gladcolor/LLM-Find. The source code of the QGIS plugin can be found at https://github.com/Teakinboyewa/AutonomousGIS_GeodataRetrieverAgent. The QGIS plugin can be downloaded and installed following the plugin User Manual.

**Appendix 1. An example prompt for selecting the data source**

Your role: A professional Python programmer in geographic information science (GIScience). You have worked on GIScience for more than 20 years and know every detail and pitfall when collecting data and coding. You know which websites you can get suitable spatial data and know the methods or tricks to download data, such as OpenStreetMap, Census Bureau, or various APIs. You are also experienced in processing the downloaded data, including saving them in suitable formats, map projections, and creating detailed and useful meta-data.

Your mission: select a suitable data source from the given list to download the requested geospatial data for this task: 1. Download all province boundaries of China mainland.
2. Save the downloaded data as polygons in GeoPackage format at: E:\China_mainland_Province_boundary.gpkg

Requirements:
1. Return the exact name of the data source as the given names.
2. If a data source is given in the task, e.g., OpenStreetMap or Census Bureau, you need to select that given data source.
3. If you need to download the administrative boundary of a place without mentioning the data sources, you can get data from OpenStreetMap.If you need to download the US Census tract and block group boundaries, download them from Census Bureau.Follow the given JSON format.
4. If you cannot find a suitable data source in the given sources, return a data source you think is most appropriate.
5. DO NOT make fake data source. If you cannot find any suitable data source, return 'Unknown' as for the 'Selected data source' key in the reply JSON format. DO NOT use ```json and ```

Data sources:
1. OpenStreetMap. You can download the administrative boundaries, street networks, points of interest (POIs) from OpenStreetMap.
2. US Census Bureau boundary. It provides the US administrative boundaries (nation, state, county, tract, and block group level, as well as metropolitan statistic areas.
3. US Census Bureau demography. It provides the demographic and socio-economic data, such as population, gender, income, education, and race.
4. US COVID-19 data by New York Times. Cumulative counts of COVID-19 cases and deaths in the United States, at the state and county level, over time from 2020-01-21 to 2023-03-23.
5. OpenWeather data. It provides historical, current, and forecast weather data. The historical data can be back to 2023-08. API limited: [Hourly forecast: 4 days, Daily forecast: 16 days, 3 hour forecast: 5 days]
6. ESRI World Imagery (for export). It is a web map service, providing satellite image tiles. You can download tiles and mosaic them into a large image.

Your reply example: {'Explanation': "According to the use requests of US state administrative boundary from OpenStreetMap, I should download data from OpenStreetMap.", "Selected data source": 'OpenStreetMap'}

**Appendix 2. An example prompt for fetching administrative boundary data**

Your role: A professional Python programmer in geographic information science (GIScience). You have worked on GIScience for more than 20 years and know every detail and pitfall when collecting data and coding. You know which websites you can get suitable spatial data and know the methods or tricks to download data, such as OpenStreetMap, Census Bureau, or various APIs. You are also experienced in processing the downloaded data, including saving them in suitable formats, map projections, and creating detailed and useful meta-data. When downloading geospatial data, the technical handbook for a particular data source is provided; you can follow it, and write Python code carefully to download the data.

Your mission: download geospatial data from the given data source for this task: 1. Download all province boundaries of China mainland.
2. Save the downloaded data as polygons in GeoPackage format at: E:\dwonloaded_data.gpkg

Data source:OpenStreetMap
Your reply example:
```python
import geopandas as gpd
import osmnx as ox
def download_data():
    # data downloading code
    # downloaded code
download_data()
```

Technical handbook:
1. If the requested area is given in an English name, you need to use `['name:en'='XX']` to filter the place in Overpass queries; otherwise you will get empty results. The `name` tag in OpenStreetMap usually is in the location language.
2. If you need to download the administrative boundary of a place from OpenStreetMap, please use a Python package named 'OSMnx' by this code line: `ox.geocode_to_gdf(query, which_result=None, by_osmid=False, buffer_dist=None)`. This method is fast.
3. If you need to download POIs, you may use the Overpass API, which is faster than OSMnx library. Code example is: `area['SO3166-2'='US-PA']->.searchArea;(nwr[amenity='hospital'](area.searchArea););out center;`
4. If you need to download polylines, you may use the Overpass API, which is faster than OSMnx library.
5. If you need to use a boundary to filter features in GeoPandas, this is the code: `gpd.sjoin(gdf, boundary, how='inner', op='within')`.
6. If you need to download multiple administrative boundaries at the same level, e.g., states or provinces, DO NOT use OSMnx because it is slow. You can use Overpass API. Example code: `area['ISO3166-1'='US'][admin_level=2]->.us;(relation(area.us)['admin_level'='4'];);out geom;`. Overpass API is quicker and simpler; you only need to carefully set up the administrative level.
7. Only use OSMnx to obtain the place boundaries; do no use it to download networks or POIs as it is very slow! Instead, use Overpass Query (endpoint: https://overpass-api.de/api/interpreter).
8. If using Overpass API, you need to output the geometry, i.e., using `out geom;` in the query. The geometry can be accessed by `returned_json['elements']['geometry']`; the gemotry is a list of points as `{'lat': 30.5, 'lon': 114.2}`.
9. Use GeoPandas, rather than OSGEO package, to create vectors.
10. If the file saving format is not given in the tasks, save the downloaded files into GeoPackage format.
11. You need to create Python code to download and save the data. Another program will execute your code directly.
12. Put your reply into a Python code block, explanation or conversation can be Python comments at the beginning of the code block(enclosed by ```python and ```).
13. The download code is only in a function named 'download_data()'. The last line is to execute this function.
14. When downloading OSM data, no need to use 'building' tags if it is not asked for.
15. You need to keep most attributes of the downloaded data, such as place name, street name, road type, and level.
16. Throw an error if the the program fails to download the data; no need to handle the exceptions.
17. If you need to convert the OpenStreetMap returned JSON to GeoJSON, you can add this line to the OverPass query: `item ::=::,::geom=geom(),_osm_type=type(), ::id=id();`. Note the converted GeoJSON may only contains polygons, no polygons.
18. This is a program for your reference; note that you can improve it:
# Below is a program to download the province boundaries of Cuba.
import geopandas as gpd
import pandas as pd
import requests
import json
from shapely.ops import linemerge, unary_union, polygonize
from shapely.geometry import MultiLineString, Polygon, MultiPolygon, LineString
from shapely.ops import polygonize

```python
def download_data():
    # Define Overpass API query to download province boundaries of Cuba
    overpass_url = "https://overpass-api.de/api/interpreter"
    overpass_query = """
    [out:json];
    area["ISO3166-1"="CU"][admin_level=2]->.cu;
    relation(area.cu)["admin_level"="4"];
    out geom;
    """

    # Send request to Overpass API
    response = requests.get(overpass_url, params={'data': overpass_query})
    response.raise_for_status()  # Automatically raises an error for bad status codes
    data = response.json()

    # Parse the JSON response
    property_list = []
    geometry_list = []

    for element in data['elements']:  # each province
        way_list = []
        outer_lines = []
        inner_lines = []

        for member in element.get('members', []):  # each way/polyline

            if 'geometry' in member:
                if member['type'] == 'way':
                    way_points = [(point['lon'], point['lat']) for point in member['geometry']]
                    line_string = LineString(way_points)
                    if member['role'] == 'outer':
                        outer_lines.append(line_string)

                    if member['role'] == 'inner':
                        inner_lines.append(line_string)

        # Create polygon. We use Multi-polygon to represent all polygons
        merged = linemerge([*outer_lines]) # merge LineStrings
        borders = unary_union(merged) # linestrings to a MultiLineString
        outer_polygons = list(polygonize(borders))
        outer_polyon = MultiPolygon(outer_polygons)

        if len(inner_lines) > 0:
            merged = linemerge([*inner_lines]) # merge LineStrings
            borders = unary_union(merged) # linestrings to a MultiLineString
            inner_polyons = list(polygonize(borders))
            inner_polyon = MultiPolygon(inner_polyons)
            final_polygon = outer_polyon.difference(inner_polyon)

        else:
            final_polygon = outer_polyon

        geometry_list.append(final_polygon)

        # extract the properties
        properties = {
```

```
            key: ', '.join(map(str, value)) if isinstance(value, list) else str(value)
            for key, value in element.items() if key not in {'geometry', 'members'}
        }
        property_list.append(properties)

    df = pd.DataFrame.from_dict(property_list)

    gdf = gpd.GeoDataFrame(df, geometry=geometry_list)
    gdf.crs = 'EPSG:4326'

    # # Save to GeoPackage
    output_file = r"E:\Cuba_Province_boundary.gpkg"
    gdf.to_file(output_file, layer='province_boundaries', driver='GPKG')

download_data()
```